\documentclass{aa}
\usepackage{graphicx}
\usepackage{txfonts}
\usepackage{natbib,twoopt}
\usepackage[breaklinks=true, colorlinks=true,urlcolor=blue,citecolor=blue,linkcolor=blue]{hyperref} 
\bibpunct{(}{)}{;}{a}{}{,}             
\makeatletter
  \newcommandtwoopt{\citeads}[3][][]{\href{http://adsabs.harvard.edu/abs/#3}%
    {\def\hyper@linkstart##1##2{}%
     \let\hyper@linkend\@empty\citealp[#1][#2]{#3}}}
  \newcommandtwoopt{\citepads}[3][][]{\href{http://adsabs.harvard.edu/abs/#3}%
    {\def\hyper@linkstart##1##2{}%
     \let\hyper@linkend\@empty\citep[#1][#2]{#3}}}
  \newcommandtwoopt{\citetads}[3][][]{\href{http://adsabs.harvard.edu/abs/#3}%
    {\def\hyper@linkstart##1##2{}%
     \let\hyper@linkend\@empty\citet[#1][#2]{#3}}}
  \newcommandtwoopt{\citeyearads}[3][][]%
    {\href{http://adsabs.harvard.edu/abs/#3}
    {\def\hyper@linkstart##1##2{}%
     \let\hyper@linkend\@empty\citeyear[#1][#2]{#3}}}
\makeatother

\begin{document}

\title{Effects of spatial resolution on inferences of atmospheric quantities from simulations}

\author{Thore E. Moe
	\inst{1,2}
	\and Tiago M.D. Pereira
	\inst{1,2}
	\and Mats Carlsson
	\inst{1,2}
	}
\institute{Rosseland Centre for Solar Physics, University of Oslo, P.O. Box 1029 Blindern, NO--0315 Oslo, Norway
\and
Institute of Theoretical Astrophysics, University of Oslo, P.O. Box 1029 Blindern, NO--0315 Oslo, Norway}
\date{}

\abstract 
{Small scale processes are thought to be important for the dynamics of the solar atmosphere. While numerical resolution fundamentally limits their inclusion in MHD simulations, real observations at the same nominal resolution should still contain imprints of sub-resolution effects. This means that the synthetic observables from a simulation of given resolution might not be directly comparable to real observables at the same resolution.  It is thus of interest to investigate how inferences based on synthetic spectra from simulations with different numerical resolutions compare, and whether these differences persist after the spectra have been spatially degraded to a common resolution} 
{We aim to compare synthetic spectra obtained from realistic 3D radiative magnetohydrodynamic (rMHD) simulations run at different numerical resolutions from the same initial atmosphere, using very simple methods for inferring line-of-sight velocities and magnetic fields. Additionally we examine how the differing spatial resolution impacts the results retrieved from the STiC inversion code.} 
{We use the RH 1.5D code to synthesize the photospheric \ion{Fe}{i}~617.33 line in local thermodynamic equilibrium (LTE), and the chromospheric  \ion{Ca}{ii}~854.209 line in non-LTE from three MHD simulation snapshots of differing spatial resolution. The simulations are produced by the Bifrost code, using horizontal grid spacing of 6 km, 12 km and 23 km, respectively. They are started from the exact same atmosphere, and the snapshots are taken after the same exact elapsed time. The spectra obtained from the high-resolution snapshots are spatially degraded to match the lowest resolution. Simple methods, like the center-of-gravity approach and the weak-field-approximation, are then used to estimate line-of-sight velocities and magnetic fields for the three cases after degradation. Finally, the spectra are input into the STiC inversion code and the retrieved line-of-sight velocities and magnetic field strengths, as well as the temperatures, from the inversions are compared.} 
{We find that while the simple inferences for all three simulations reveal the same large-scale tendencies,  the higher resolutions yield more fine-grained structures and more extreme line-of-sight velocities/magnetic fields in concentrated spots even after spatial smearing. We also see indications that the imprints of sub-resolution effects on the degraded spectra result in systematic errors in the inversions, and that these errors increase with the amount of sub-resolution effects included. Fortunately, however, we find that including successively more sub-resolution yields smaller additional effects; i.e. there is a clear trend of diminishing importance for progressively finer sub-resolution effects.} 
{} 


\maketitle

\section{Introduction}
\label{sec:intro}

\begin{figure*}
	\centering
		\includegraphics[width=18cm]{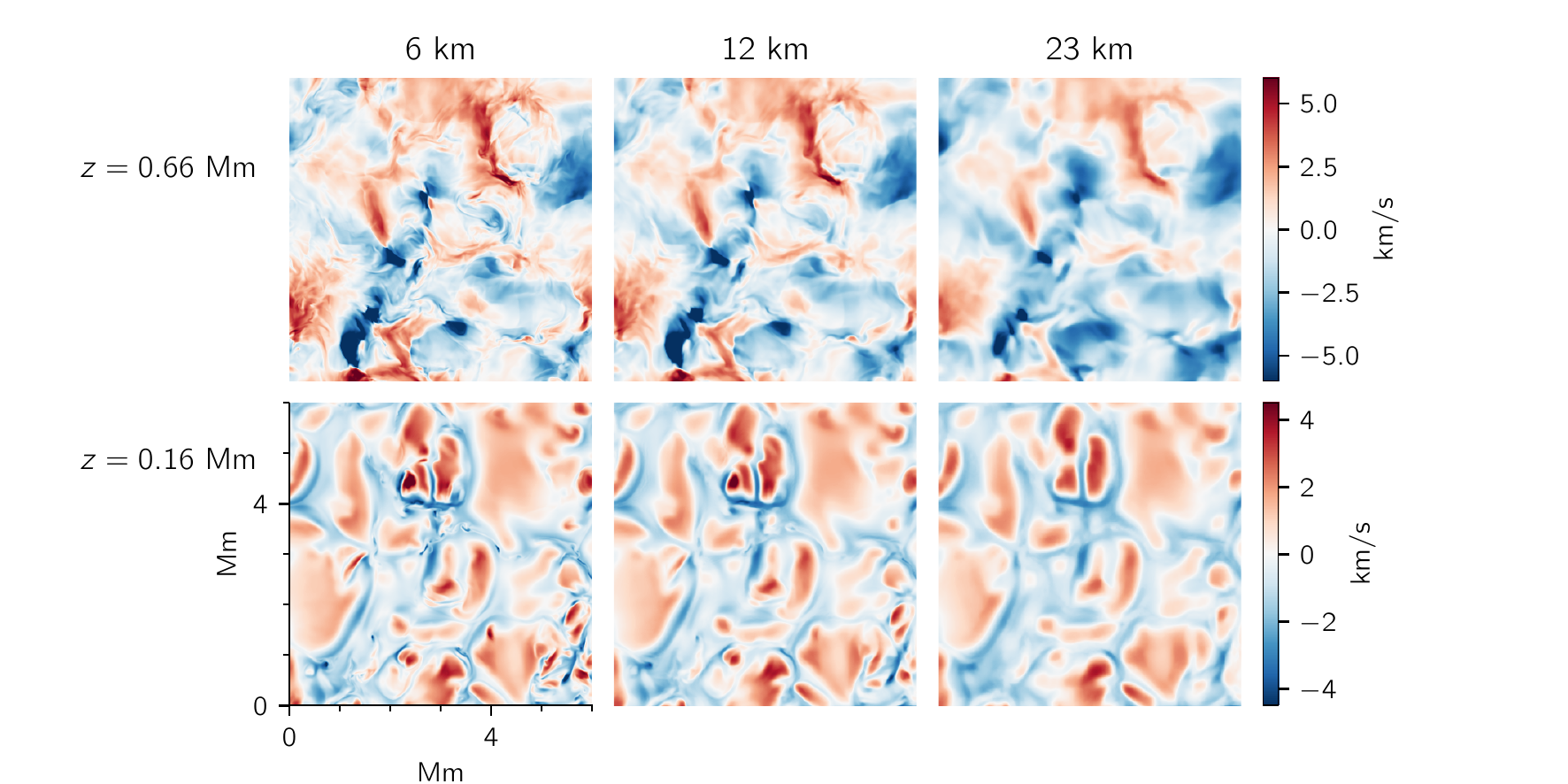}
			\caption{Line-of-sight velocities taken directly from horizontal slices of the simulations at different heights. The left column shows the 6 km simulation, the middle column shows the 12 km simulation and the right column shows the 23 km simulation. The top row corresponds to a height of 0.66 Mm, i.e. in the low chromosphere, while the bottom row corresponds to a height of 0.16 Mm, i.e. in the photosphere. Note that we here, and in all subsequent colormaps, only show a quarter of the full simulation extent, in order to more easily see fine details.}
			\label{fig:sim_vel}
\end{figure*}

Small-scale magnetic fields are believed to be strongly linked to the dynamics and heating of the solar atmosphere (see e.g. \citeads{2009SSRv..144..275D}, \citeads{2009ApJ...700.1391M}, \citeads{2010ApJ...723L.134B}, \citeads{2020ApJ...894L..17Y}). Today, 3D radiative magnetohydrodynamic (rMHD) simulations are an increasingly important tool to study the evolution and impact of solar magnetic fields \citepads{2012LRSP....9....4S}. These simulations are necessarily limited by computational constraints in how finely they can describe the atmosphere. That is, the numerical grid spacing used presents a limit for how small spatial features can be included. Observations also have finite spatial resolution, but they will still contain imprints of sub-resolution effects. Small-scale motions still happen in the atmosphere, even if they are not resolved by our telescopes. This is in contrast to simulations, where a voxel really does represent the smallest details available. It is therefore difficult to directly compare simulations and observations at the same nominal resolution. In most studies (e.g. \citeads{2005A&A...442.1059K}, \citeads{2009A&A...507..417P}, \citeads{2010A&A...513A...1D}, \citeads{2021A&A...647A..46S}) synthetic spectra from high-resolution numerical simulations are spatially (and spectrally) degraded using convolutions with point spread functions (PSFs) and downsampled to match observational resolutions. In such studies the discussion of resolution effects is limited to the mapping between the simulation and the observations. The questions we address here are of a different nature. They are focused on the effects of inferring physical quantities from spectra captured at lower resolutions. In the absence of in-situ measurements, spectra remain the primary way to infer physical quantities in stellar atmospheres. The accuracy of such inferrals, whether performed directly from spectra (e.g. Doppler shifts) or via inversions, depends critically on the parameters of the observation, such as spectral and spatial resolution, stray-light, and other instrumental effects. The complex problem of stray-light can also be seen as a spatial degradation, and many ways to mitigate it have been proposed over several decades (e.g. \citeads{1987ApJ...322..473S}, \citeads{2007PASJ...59S.837O}, \citeads{2007ApJ...655..615L}, \citeads{2012A&A...548A...5V}). On the other hand, the issue of sub-resolution effects has received scant attention.

The key question we want to answer is: what are the effects of sub-resolution on inferred quantities? Spectra will carry subtle signatures of sub-resolution, which presumably could affect inferred quantities such as line-of-sight velocities or magnetic fields. This cannot be tested with observations alone, since they already contain an imprint of all scales taking place on the solar surface. But it can be studied with synthetic spectra from simulations, which can be generated with different amounts of ``sub-resolution structure''. Our question is relevant both from a modelling perspective as well as an observational perspective. From the modelling side, it would be most useful to know how much numerical resolution is enough to study events observed at a given resolution. From the observational side, it would be useful to know how much sub-resolution effects influence quantities inferred from spectra observed at a given resolution. 
Both for inferring quantities through direct analysis of spectra, or via inversions, where typically each observed pixel is assumed to be an independent, plane-parallel atmosphere with quantities varying smoothly with height \citepads[see review by][]{2017SSRv..210..109D}.
 
\section{Methods}
\label{sec:methods}

\subsection{Overall Approach}

Our approach can be summarized as (1) generating synthetic spectra with different amounts of substructure, (2) inferring physical quantities from them, and (3) analyzing the differences. In this section we detail how we achieve the first two points, and proceed with the analysis of the results in Section~\ref{sec:results}. 

How does one create spectra with varying amounts of sub-pixel structure? Our basic premise is to generate synthetic spectra from 3D rMHD simulations, where all spectra share the same spatial pixel size. Spectra with no sub-pixel structure are obtained by direct synthesis from a simulation. In this case, a spatial pixel from the simulation has the same resolution as the spatial pixel of the synthetic spectra. Spectra with sub-pixel structure are obtained by synthesis from a higher resolution simulation, followed by spatial degradation and rebinning to a lower resolution pixel size (to mimic what happens with observed light). As detailed below, our reference ``zero sub-pixel structure'' simulation has a horizontal spatial resolution of 23~$\mathrm{km}\,\mathrm{pix}^{-1}$. We then work with simulations that have two times and four times higher resolutions (approximately 6 and 12~$\mathrm{km}\,\mathrm{pix}^{-1}$). Spectra from the higher resolution simulations are then spatially degraded so they match the spatial resolution of the canonical simulation, and therefore have different amounts of sub-resolution effects. The 6~$\mathrm{km}\,\mathrm{pix}^{-1}$ run has imprints of resolution effects up to four times the canonical resolution, and the 12~$\mathrm{km}\,\mathrm{pix}^{-1}$ run has up to twice the canonical resolution.

In the rest of this section we give details on the simulations used, how the spectral synthesis was performed, and how physical quantities were inferred from the spectral quantities.

\subsection{Simulations}
\label{subsec:simulations}

\begin{figure*}
\centering
	\includegraphics[width=18cm]{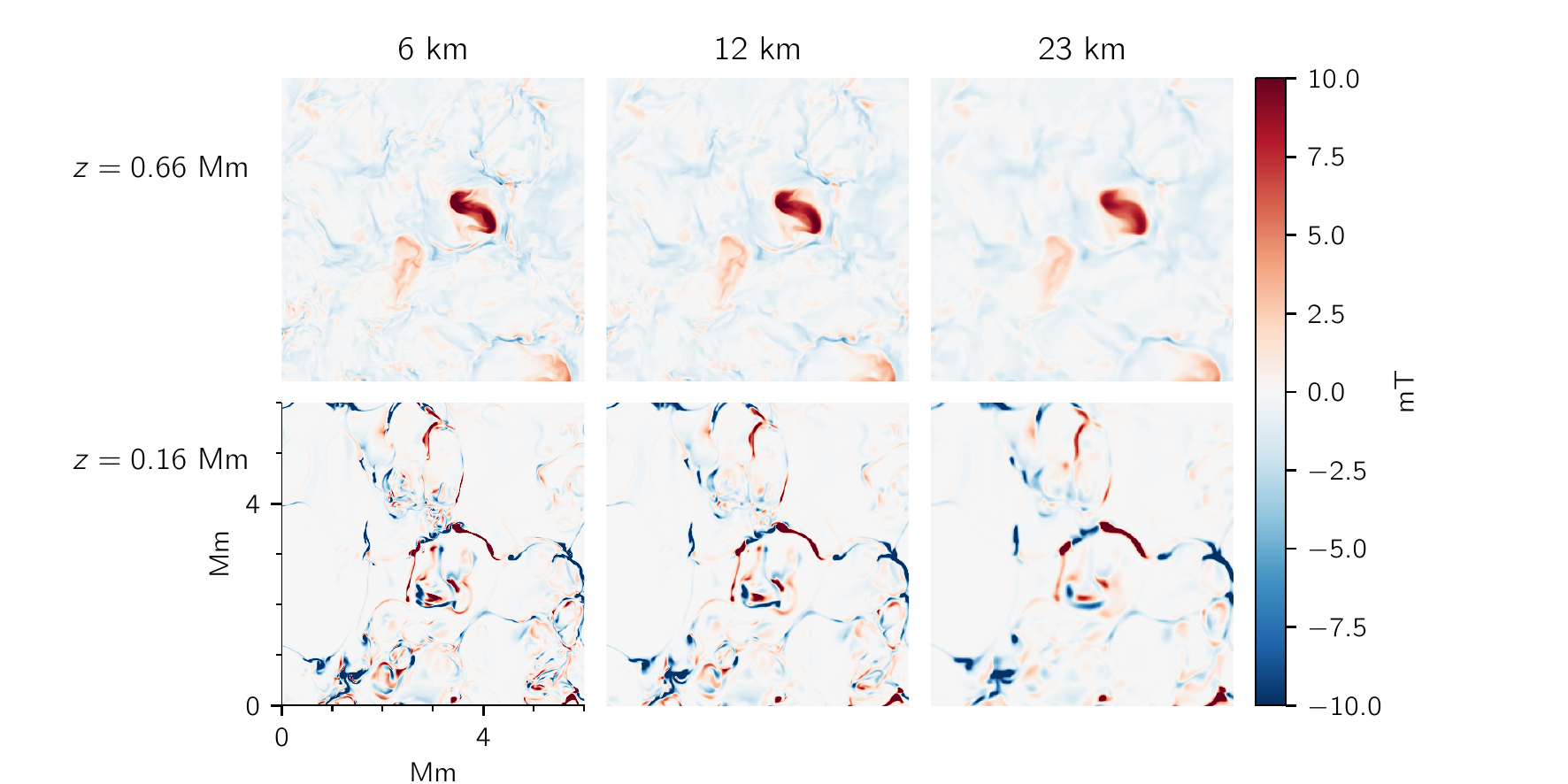}
		\caption{Line-of-sight magnetic field taken directly from horizontal slices of the simulations at different heights. The left column is from the 6 km simulation, the middle column is from the 12 km simulation and the right column is from the 23 km simulation.The top row shows slices at a height of 0.66 Mm, i.e. in the low chromosphere, while the bottom row shows slices at a height of 0.16 Mm, i.e. in the photosphere.}
		\label{fig:sim_mag}
\end{figure*}

We make use of three same-time rMHD simulation snapshots with a magnetic field configuration similar to that of a coronal hole, identical in all respects apart from their spatial resolution. The simulations were run using the Bifrost code \citepads{2011A&A...531A.154G}. The challenge was to develop simulations whose state and morphology are as close as possible, with the only difference being the spatial resolution. Our approach was as follows. We started by running a simulation with $1024\times1024\times1024$ grid points, with a horizontal resolution of 12~$\mathrm{km}\,\mathrm{pix}^{-1}$ and vertical resolution that varied (non-equidistant). This simulation was allowed to evolve for about 2.2~h of solar time. At this point, two other simulation were branched off and run in parallel. One of the additional simulations was run with exactly half the number of grid points in each direction ($512\times512\times512$ cells with 23~$\mathrm{km}\,\mathrm{pix}^{-1}$ in the horizontal direction), and the other was run with twice as many points in the horizontal directions ($2048\times2048\times1024$ cells with 6~$\mathrm{km}\,\mathrm{pix}^{-1}$ in the horizontal direction). The spatial extent and all other simulation parameters were the same in all simulations. All simulations were allowed to evolve for a further 120~s of solar time, and those were the snapshots we used for this work. This time was chosen to balance two opposing considerations: first the simulations should have had sufficient time to relax so that the high resolution structures have coalesced at their appropriate resolution level, secondly the duration should not be so long that the simulations diverge from each other in a macroscopic sense. Inspection of the spatial power spectra for the continuum intensity shows that the power at the smallest scales has just about stabilized at the chosen time, while the power at the large scales remains virtually identical. This gives us some confidence that the chosen snapshots reasonably balance our competing requirements.

As noted before, we will use the 23~km resolution simulation as a reference for no spatial sub-resolution, and the other two for varying amounts of sub-resolution. This means that we have in effect only two data points other than the reference. One could argue that this is insufficient, that one could still experiment with a broader range in resolutions. However, there are also good reasons to keep it at three. First, it is numerically unstable to branch off a simulation with a large difference in spatial resolution compared to the parent simulation. Such simulations would need to be relaxed for longer, and the result would be a simulation where the morphology would have evolved differently from the parent simulation, making it much harder to compare features at the pixel level. Second, since we are degrading the higher resolution simulations to a lower resolution, there will be diminishing returns after a certain resolution. The sub-resolution imprints on spectra would be smaller and smaller because higher resolution details would be destroyed in the spatial degradation process. For these reasons we chose to use three simulations.

The simulations cover approximately 12 Mm in the horizontal directions, and extend from $z=-2.5$~Mm below and up to $z=8$~Mm above the solar surface ($z=0$ is defined where average $\tau_{500}=1$). The vertical grid points are more densely distributed around the chromospheric and photospheric heights, tapering off towards the corona above and the convection zone below (6--8~km spacing in the height range $-1$ to 3~Mm and increasing to 35~km at the top and 15~km at the bottom of the simulation domain in the high resolution run). The simulations were run with periodic boundary conditions along the horizontal dimensions, and open boundary conditions for the vertical dimension. In all runs we assume hydrogen ionization in local thermodynamic equilibrium (LTE). The magnetic field configuration comes from a small-scale dynamo, with an additionally imposed vertical field  having a mean signed $B_z$ of 0.25 mT (2.5 G) and a mean unsigned magnetic field of 3.7 mT at $z=0$.

Lastly, a note on spatial resolution and numerical pixel size. With observations, the spatial resolution comes from the diffraction limit of the telescope, which gives the smallest detail observable. With numerical simulations, the relation between pixel size and smallest detail that can be reproduced is not so clear cut because of numerical diffusion. \citetads{2011ApJ...737...13K} find that numerical dissipation in the STAGGER code (a close relative of Bifrost) leads to considerable damping of scales $\lesssim 5$ grid cells. This means that numerical resolution cannot be directly compared to observational resolution, since numerical diffusion makes it unlikely the code will produce small scales down to the grid resolution. For example, in Figure~\ref{fig:sim_vel} we show vertical velocities at two heights from the simulations. Looking at sizes of the smallest structures, it is clear they are larger than the pixel sizes. To estimate the real resolution of the 23~km simulation, we performed a simple test. We extracted a high-resolution map of vertical velocities at $z=0$ from the 6~km simulation (resampled to 23~km pixels) and degraded it to find out how much degradation best agrees with the corresponding map from the 23~km simulation. That is, we take the high-resolution map as a ``ground-truth'', and estimate the impact of numerical diffusion as the amount of degradation required until the degraded ``ground-truth'' matches the map from the 23~km simulation. We find that for a Gaussian convolution, a FWHM of about 5 pixels works best, which is consistent with the results of \citetads{2011ApJ...737...13K} for the STAGGER code. This means that the ``smallest detail observable'', or equivalent observational resolution, of the 23~km simulation is $\approx5$ times the pixel size, or about 115~km.

\subsection{Spectral synthesis}
\label{subsec:synthesis}

To probe the effects of sub-resolution in the photosphere and in the chromosphere, we synthesize two spectral lines: \ion{Fe}{i}~617.33 nm (computed in LTE), and  \ion{Ca}{ii}~854.209 nm (computed in non-LTE, or NLTE). Note that we use air wavelengths throughout this work. The \ion{Fe}{i} line is formed in the photosphere and is a commonly-used diagnostic for photospheric magnetic fields; in particular it is used by the HMI-instrument  \citepads{2012SoPh..275..229S}. The \ion{Ca}{ii}~854.209 nm line is formed over a large range of heights, from the photosphere to the chromosphere, and suffers from strong NLTE effects \citepads{2009ApJ...694L.128L}. There is considerable interest in using this line and its neighbouring relatives which together constitute the \ion{Ca}{ii} infrared triplet, for diagnostics of the chromosphere, (see e.g. \citeads{2017SSRv..210..109D}, \citeads{2016MNRAS.459.3363Q,2017MNRAS.464.4534Q}).

Both lines were synthesized using the RH 1.5D code \citepads{2015A&A...574A...3P}, a publicly-available line synthesis code based on the RH code \citepads{2001ApJ...557..389U}. RH 1.5D works on a column-by-column basis, solving the radiative transfer equation treating each column in the 3D simulation as a 1D atmosphere. To save computing time, points above 100 kK in the top of each 1D atmospheric column were not included in the calculation. Line profiles for both lines were computed assuming complete redistribution (CRD). For the NLTE computations of \ion{Ca}{ii} we used a model atom with five bound levels plus a continuum level. For both lines, we solved the polarized radiative transfer equation assuming only Zeeman polarization.

The choice of using LTE for the synthesis of \ion{Fe}{i}~617.33 nm deserves some further comment. While it has been usual to assume LTE for this line, and the very similar \ion{Fe}{i} 630.15/630.25 nm pair, it has long been known that both NLTE-effects and 3D-effects can have an impact on these photospheric \ion{Fe}{I} lines (see e.g. \citeads{2013A&A...558A..20H,2020A&A...633A.157S,2021A&A...647A..46S} for recent investigations on this subject). In this work, however, we only aim to compare spectra across different spatial resolutions; we are not \textit{per se} aiming to retrieve the most faithful estimations of the atmospheric parameters possible. Furthermore, the wings of \ion{Ca}{ii}, which \textit{are} computed in NLTE, also probe the photosphere and, as we will show, display the same tendencies for the inferred atmospheric parameters as the LTE \ion{Fe}{i} line. We therefore believe the LTE computations of the \ion{Fe}{i}-line to be sufficient for our purposes.

The last step was to spatially degrade the spectra from the 6 km and 12 km simulations, to a resolution of 23 km so they could be directly compared with the simulation run at 23 km (the reference simulation with no sub-pixel structure). For each wavelength in the synthetic profiles, we performed a 2D Gaussian convolution in the spatial domain. The FWHM was 4 pixels for the 6 km simulation and 2 pixels for the 12 km resolution. Lastly, the spectrograms were downsampled to a 23 km pixel size. The end result were spectrograms with the same size as those computed from the 23 km simulation ($512\times512$ columns). No analysis was performed for the 6 and 12 km resolution spectra at their native resolutions. In all subsequent mentions, when we refer to results from 6 km or 12 km, we mean $6\to23$~km or $12\to23$~km: the simulations degraded to a pixel size of 23 km.

\subsection{Inferring physical quantities from spectra}
\label{subsec:inference}

Equipped with synthetic spectra with different amounts of sub-resolution effects, we proceeded to infer physical quantities from them. We start with widely-used techniques of deriving line-of-sight velocities and magnetic fields from spectra. Those are detailed below, with different strategies for photospheric and chromospheric values.

\subsubsection{Photosphere}
\label{subsubsec:photosphere}

To infer the photospheric line-of-sight velocities and magnetic fields, we make use of the \ion{Fe}{i}~617.33 nm line. We apply the so-called center-of-gravity-method (COG-method), following \citetads{2003ApJ...592.1225U}. The central wavelength of the line is estimated for each pixel as:

\begin{equation}
\label{eq:cog_lambda}
\lambda_{COG} = \frac{\int \lambda (I_{cont} - I(\lambda)) d\lambda}{\int (I_{cont} - I(\lambda)) d\lambda},
\end{equation}
where $\lambda_{COG}$ is the estimated central wavelength, $I_{cont}$ is the Stokes I intensity of the nearby continuum, $I(\lambda)$ is the wavelength-dependent Stokes I intensity and the integral extends over the whole line profile. (Note that SI units are used throughout this work.) In essence, this is a mean wavelength of the profile weighted by the line intensity's departure from the continuum. The position of the central wavelength is then converted to line-of-sight velocity using the Doppler formula, using the convention where positive velocities are upflows. 

The line-of-sight magnetic fields can be estimated by
\begin{equation}
\label{eq:blos_cog}
B_{LOS} = \frac{\lambda_+ - \lambda_-}{2} \frac{4\pi m_e c}{e g_L \lambda_0^2},
\end{equation}
where $B_{LOS}$ is the line-of-sight magnetic field, $m_e$ is the electron mass, $e$ is the electron charge, and $g_L$ is the (effective) Land\'{e} g-factor of the line. $g_L$ is 2.5 for \ion{Fe}{i}~617.33 nm and 1.1 for \ion{Ca}{ii}~854.209 nm; an explicit formula for the g-factor can be found in \citetads{2004ASSL..307.....L}. $\lambda_+$ and $\lambda_-$ are the centroids of the circularly polarized line components, right- and left-handed respectively. These are obtained by replacing $I(\lambda)$ with $I(\lambda) \pm V(\lambda)$ in Eq. (\ref{eq:cog_lambda}), where $V(\lambda)$ is Stokes V. We use the convention where positive $B_{LOS}$ are pointing outward.

The same method may be applied to the \ion{Ca}{ii} line, though in that case it is not the chromospheric line core which is probed; but rather the much wider photospheric line wings, which carry most of the `weight' in the COG-method. That is because the full line profile must be integrated over, in order to avoid biasing the estimates. From a purely geometric standpoint the wings of the profile are so much wider than the core that the shape and size of the wings make them contribute much more to the center of gravity. That is why the full-profile application of the COG-method is mostly sensitive to the photospheric conditions of the line-wings. In order to increase the relative contribution of the core to the COG-estimate, one would have to limit the integration window to contain less area in the wings. But that is challenging to do without introducing a strong bias in the estimates. If, for instance, the integration is not centered on the (Doppler-shifted) line core, more signal will be included from one wing than the other, artificially shifting the COG-estimate.

\begin{figure*}
	\centering
		\includegraphics[width=18cm]{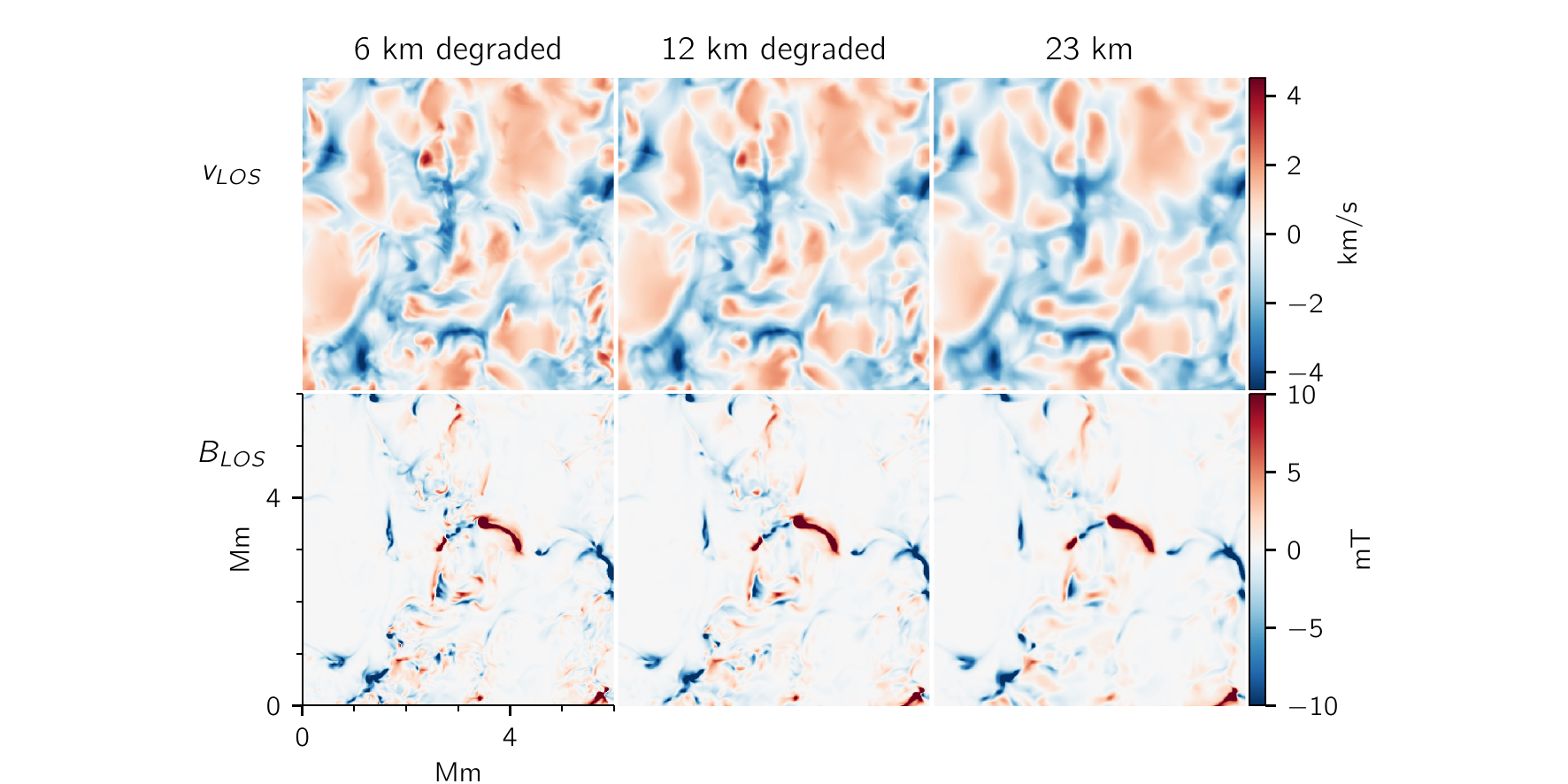}
			\caption{Inferred line-of-sight velocities and magnetic fields for \ion{Fe}{i}~617.33 nm. The top row shows the line-of-sight velocities computed by the COG-method, while the bottom row shows the line-of-sight magnetic fields calculated by the COG-method. The left column shows the results from the degraded 6 km simulation (degraded to match the resolution of the 23 km simulation), the middle-right column shows the results from the degraded 12 km simulation (degraded to match the resolution of the 23 km simulation) and the right column shows the results for the (undegraded) 23 km simulation.}
			\label{fig:fei_2x3}
\end{figure*}

\subsubsection{Chromosphere}
\label{subsubsec:chromosphere}

Since the COG-method only works well when applied to a full profile, we must use a different approach to retrieve information from the chromospheric core of \ion{Ca}{ii}. Given that we introduced no spectral degradation and the simulations used are of moderately high resolution, the \ion{Ca}{ii}~854.209 line profiles show great variability near the line core. The core itself is poorly defined, and often many peaks and minima can be found, which complicates the task of extracting a single velocity value. Usual methods such as measuring the shift of the line minimum become unreliable. Given these difficulties, and how they affect differently the simulations considered, we believe that inferring chromospheric velocities for our purposes is fraught with uncertainty, and opted not to estimate chromospheric velocities from spectra. 

For estimating the line-of-sight magnetic field strengths in the chromosphere we are able to employ the weak field approximation (WFA, see \citeads{2009ApJ...700.1391M} and  \citeads{2018ApJ...866...89C}). We use, corresponding to Eq. (8) in \citetads{2018ApJ...866...89C}, the following estimate (in SI units):
\begin{equation}
\label{eq:wfa_blos}
B_{LOS} = -\frac{4\pi m_e c}{e g_L \lambda_0^2} \frac{\Sigma_{\lambda} \frac{\partial I(\lambda)}{\partial\lambda} V(\lambda)}{\Sigma_{\lambda} (\frac{\partial I(\lambda)}{\partial\lambda})^2},
\end{equation}
where the sum is over all wavelength points in the region of interest. For the sake of simplicity, we employ the method on the interval $\lambda_0 \pm 0.05$ nm, which seems to reasonably cover the core for most of our profiles.

\subsection{Inferring physical quantities from inversions}
\label{subsec:inversions}

Our final means of comparing the effects of spatial resolution is to see how they impact the atmospheres retrieved from inversions. For each of the three cases (6 km degraded to 23 km, 12 km degraded to 23 km, 23 km undegraded) we run one inversion cycle with both the \ion{Fe}{i} and \ion{Ca}{ii} spectra as input to the state-of-the-art inversion code STiC \citepads{2019A&A...623A..74D}. STiC, as RH 1.5D, works on a column by column basis. As an initial guess atmosphere for all pixels, we use a 1D model constructed from the 23 km simulation averaged over iso-surfaces of column mass. The averaged model is constructed by taking the horizontal average of the atmosphere interpolated at 100 evenly log-spaced points in column mass in the range $[-7,1]$~kg~m$^{-2}$, and afterwards excluding all but the first point in either direction outside the interval $[-7,1]$ in $\log(\tau_{500})$ so that regions with no signal from the spectral lines are not included in the inversions. The inversions were performed with 9 nodes for temperature, no microturbulence, and 5 nodes for the line-of-sight magnetic field strength, the horizontal magnetic field strength, the azimuth angle of the magnetic field, and the line-of-sight velocity.

\section{Results}
\label{sec:results}

\subsection{Comparing the simulations}
\label{subsec:sim_results}

Figure \ref{fig:sim_vel} shows the vertical velocities $v_z$ for the three simulations at chromospheric and photospheric heights, at their native resolutions. Only a quarter of the full spatial extent is shown, to better show the small details. These two heights were not chosen to correspond directly with the formation heights of the investigated spectral lines. Rather, we want to show the general and qualitative differences between the simulations in the photosphere and chromosphere. We see that, although all three simulations paint very much the same large-scale picture, the higher-resolution simulations display fine-structure that is absent in the lower-resolution cases. Furthermore, they contain localized ``hotspots'' showing more extreme line-of-sight velocities than in the 23 km simulation. 

The same story is corroborated through Fig. \ref{fig:sim_mag}, which shows the simulations' vertical magnetic fields $B_z$ at chromospheric and photospheric heights. The differences in fine-structure are particularly noticeable in the bottom panel, but can also be seen in the top panel.  As for the vertical velocities, localized areas appear where the field strengths are more extreme in the high-resolution cases.

\subsection{Photospheric inferences from spectra}
\label{subsec:photospheric_results}
Turning our attention to the physical quantities inferred from the spectra, we first look at the information gleamed from the \ion{Fe}{i}~617.33 nm line. The three  columns of Fig. \ref{fig:fei_2x3} show the inferred line-of-sight velocities (top) and magnetic fields (bottom) for this line, using the COG-method, for the three resolution cases after spatial degradation to the same nominal resolution; i.e. 6 km degraded to 23 km, 12 km degraded to 23 km and 23 km undegraded, going from left to right. 

\begin{figure*}
	\centering
		\includegraphics[width=18cm]{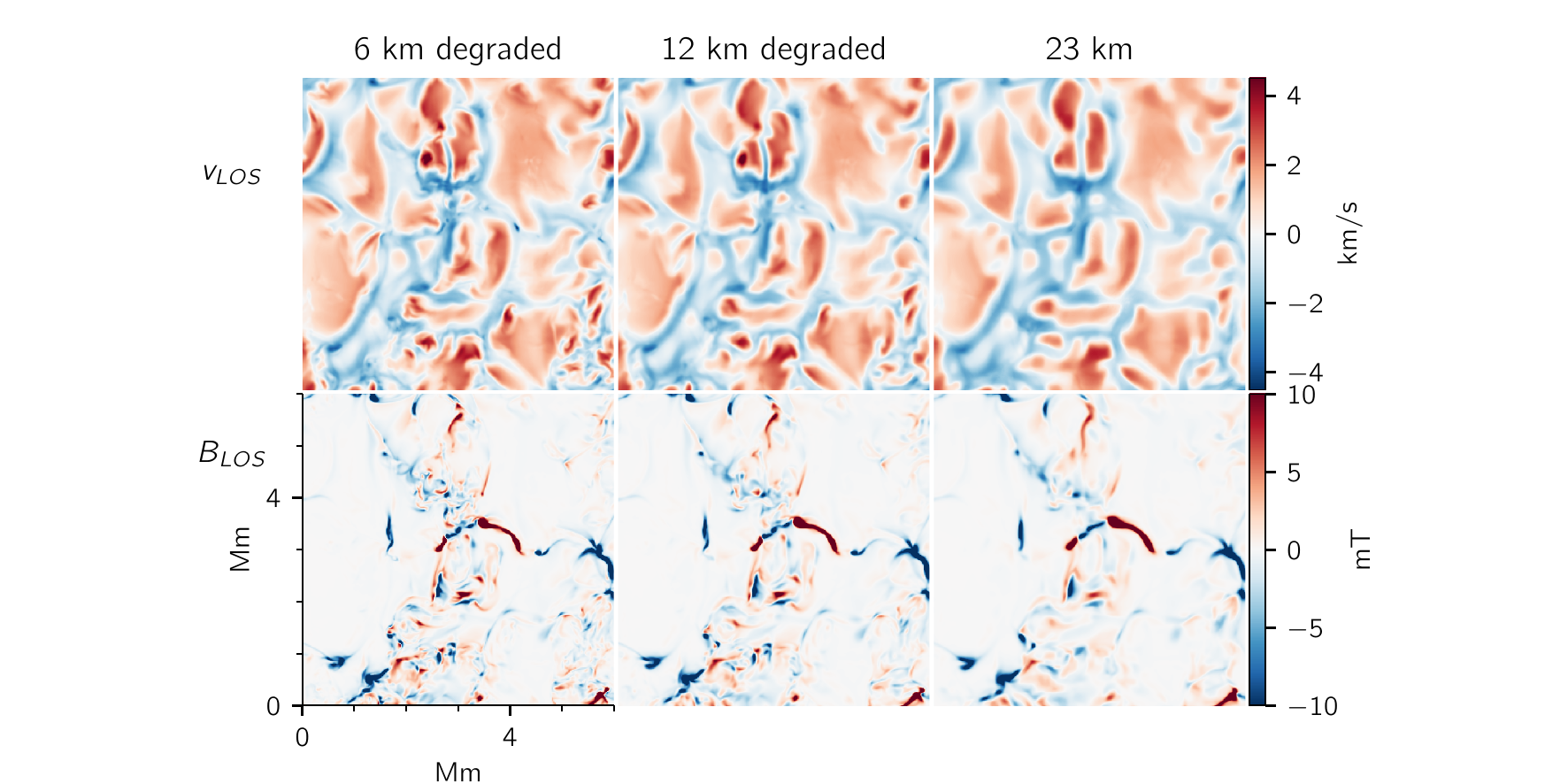}
			\caption{Inferred line-of-sight velocities and magnetic fields for \ion{Ca}{ii}~854.209 nm, using the COG-method over the whole line profile (core and wings). The layout is exactly the same as in Fig. \ref{fig:fei_2x3}.  }
			\label{fig:caii_2x3}
\end{figure*}

As seen in the quantities from cuts in the simulations, the inferred line-of-sight velocity and magnetic field strength maps continue to show the same large-scale structures. But interestingly, even after the spatial degradation to the same nominal resolution of 23 km, significant differences in fine-structure and extrema persist between the three cases. In particular, as the underlying resolution of the simulation increases, there is a tendency, in tandem, of more fine-structure and more extreme ``hotspots''. 

In Fig. \ref{fig:caii_2x3} we show the corresponding line-of-sight velocity and magnetic field maps derived from the \ion{Ca}{ii}~854.209 nm line. As before, these maps were obtained using the COG-method over the whole line profile.  As mentioned in Sect. \ref{subsubsec:photosphere}, the contribution from the line wings dominates and gives results which are clearly photospheric in nature. Though not exactly matching Fig. \ref{fig:fei_2x3}, the same features and tendencies can be observed in this case as well. In accordance with the photospheric slices of the simulations in Figs. \ref{fig:sim_mag} and \ref{fig:sim_vel}, we see increasing fine-structure and extrema with increasing resolution. While the \ion{Fe}{i} line and the \ion{Ca}{ii} wings are not formed at the same heights, the striking similarities between the line-of-sight velocities and field strengths retrieved from them and the photospheric simulation slices lends confidence to our choice of ignoring NLTE-effects in the \ion{Fe}{i}-synthesis. That is, the effects of spatial resolution appear similar in the photosphere for both LTE and NLTE synthesis, at least for the currently considered case.

In Fig. \ref{fig:photo_hist} we show histograms of both quantities for the two lines, taken for the full horizontal extent of the simulations, not just the smaller field of view shown in the previous figures. These corroborate our qualitative discussion above. The left column, showing the logarithmic distribution of the inferred line-of-sight velocities for \ion{Fe}{i} (top) and \ion{Ca}{ii} (bottom) displays the tendency of increased extremes in the line-of-sight velocity with increased resolution. While the jump from 23 km to 12 km notably seems more significant than the one from 12 km to 6 km, the line-of-sight velocities do not appear converged yet. The right column, showing the logarithmic distribution of the inferred line-of-sight magnetic field strengths, paints a more equivocal picture. It is clear that the distribution is pushed to larger extremes for the degraded 6 km and 12 km cases compared to inferences from the native 23 km simulation. Here, however, it seems that the extremities don't grow much from the 12 km to 6 km resolution. But, as we saw in Figs. \ref{fig:fei_2x3} and \ref{fig:caii_2x3}, even with comparable extreme values the fine-structure has not converged in the 12 km case.

\subsection{Chromospheric inferences from spectra}
\label{subsec:chromospheric_results}

\begin{figure*}
	\centering
		\includegraphics[width=18cm]{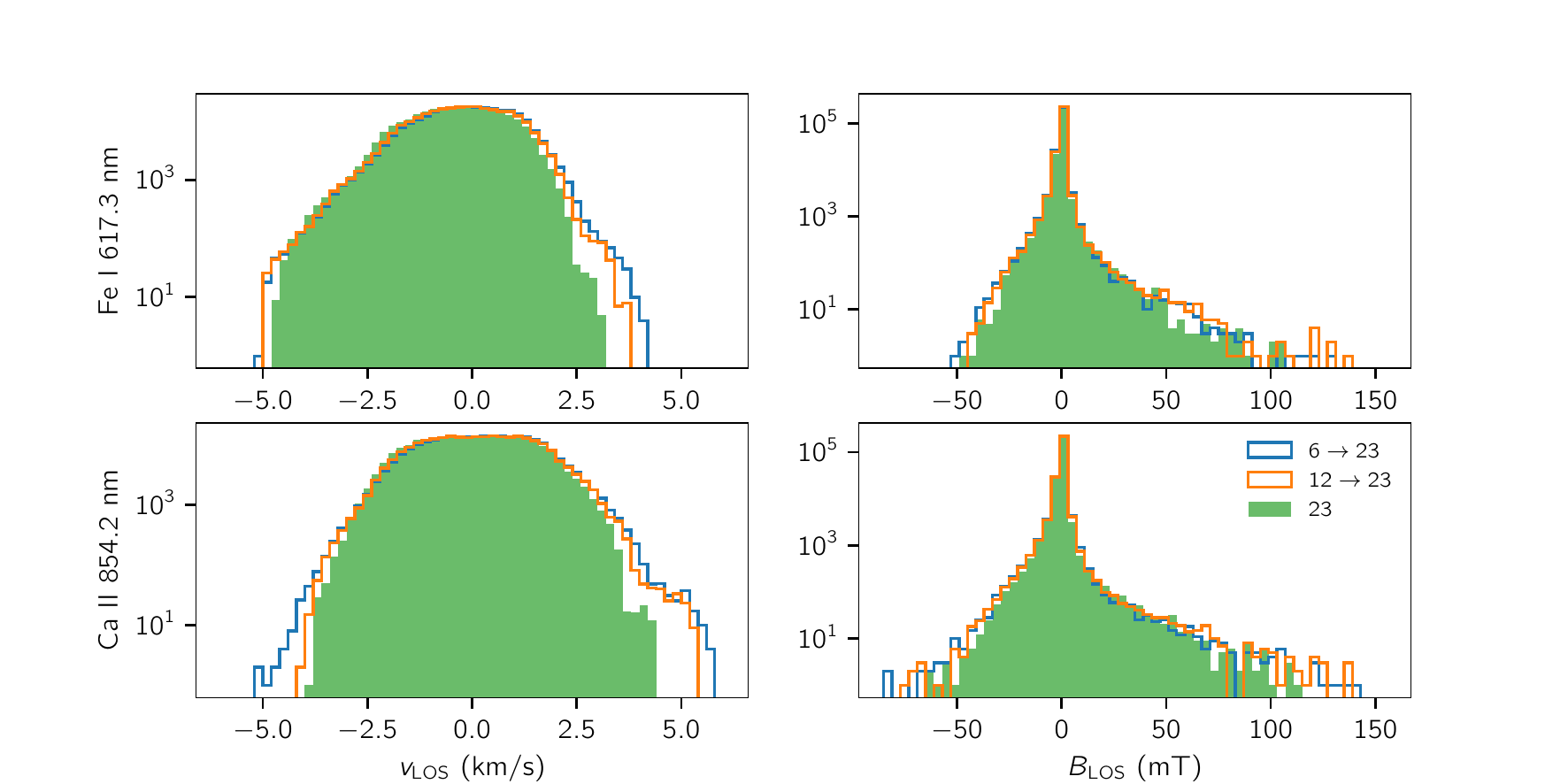}
			\caption{Histograms of the inferred quantities using the COG-method on the whole line profile for \ion{Fe}{i} 617.33 nm (top) and \ion{Ca}{ii} 854.209 nm (bottom). Note that here, as in Fig. \ref{fig:core_mag_hist}, we include the whole field of view for the simulation in the histograms, not only the quadrant shown in the colormaps elsewhere.}
			\label{fig:photo_hist}
	\end{figure*}

\begin{figure*}
	\centering
		\includegraphics[width=18cm]{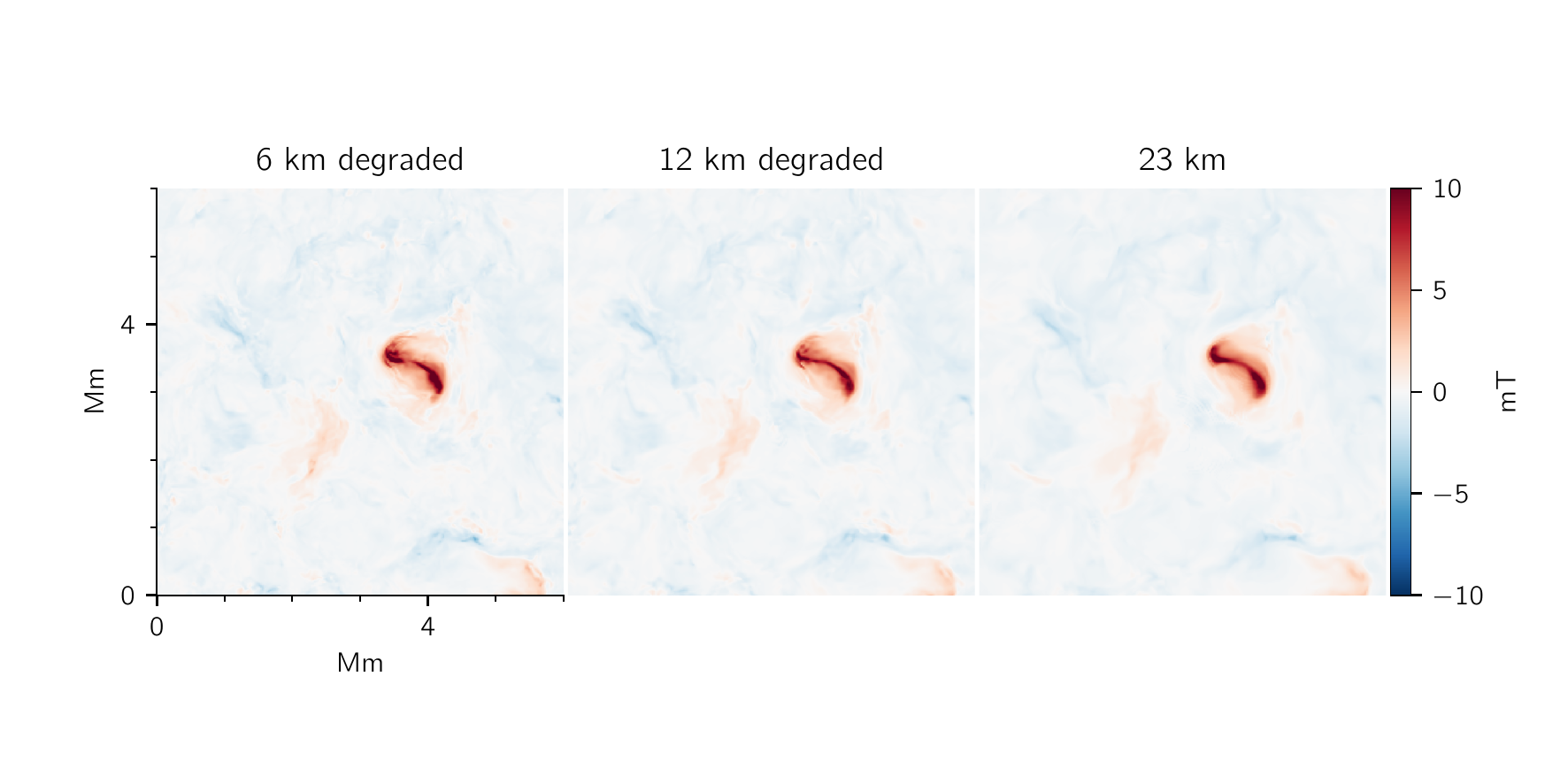}
			\caption{Inferred line-of-sight magnetic fields for the core of the \ion{Ca}{ii}~854.209 nm line. The fields are estimated from Eq. \ref{eq:wfa_blos} applied to the  interval $\pm$ 0.05 nm from the nominal center wavelength of the line. The layout is otherwise the same as in Fig. \ref{fig:fei_2x3} and Fig. \ref{fig:caii_2x3}.}
			\label{fig:core_mag}
	\end{figure*}

\begin{figure}
	\centering
		\includegraphics[width=8cm]{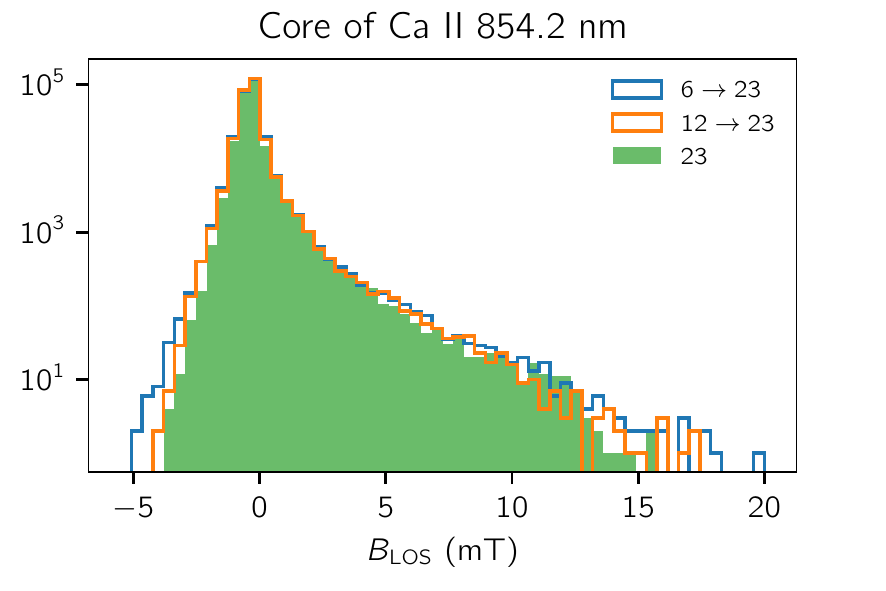}
			\caption{Histograms of the inferred line-of-sight magnetic field strengths for the \ion{Ca}{ii} core after degradation to the common 23 km resolution, using the WFA-method Eq. \ref{eq:wfa_blos}. Note that here, as in Fig. \ref{fig:photo_hist}, we include the whole horizontal extent of the simulation in the histograms, not only the quadrant shown in the colormaps elsewhere.}
			\label{fig:core_mag_hist}
	\end{figure}

As for our chromospheric inferences, Fig \ref{fig:core_mag} shows the retrieved line-of-sight magnetic field strengths using Eq. \ref{eq:wfa_blos} on the interval $\pm 0.05$ nm around the nominal line core. The differences between the resolution cases are not as dramatic here in the chromosphere as in the previously discussed photospheric estimates. Yet, looking closely at the main magnetic feature, we do see additional fine-structure and concentration of the line-of-sight magnetic field with increasing resolution. The (logarithmic) distribution of the inferred line-of-sight magnetic field strengths obtained from this method is shown in Fig. \ref{fig:core_mag_hist}; as with Fig. \ref{fig:photo_hist} this distribution also covers the whole horizontal range of the simulation, not just the highlighted quadrant in Fig. \ref{fig:core_mag}. The most salient aspect of this distribution is that, in contrast to the photospheric estimates of the line-of-sight magnetic field strength, the chromospheric inferences retain the tendency of getting more extreme from both 23 km to 12 km as well as from 12 km to 6 km in resolution. So while the most extreme values for the photospheric line-of-sight magnetic field seem to be reached already at 12 km resolution, the same is not the case for the chromospheric estimates which are still increasing in sync with the resolution.

	\begin{figure*}
	\centering
		\includegraphics[width=18cm]{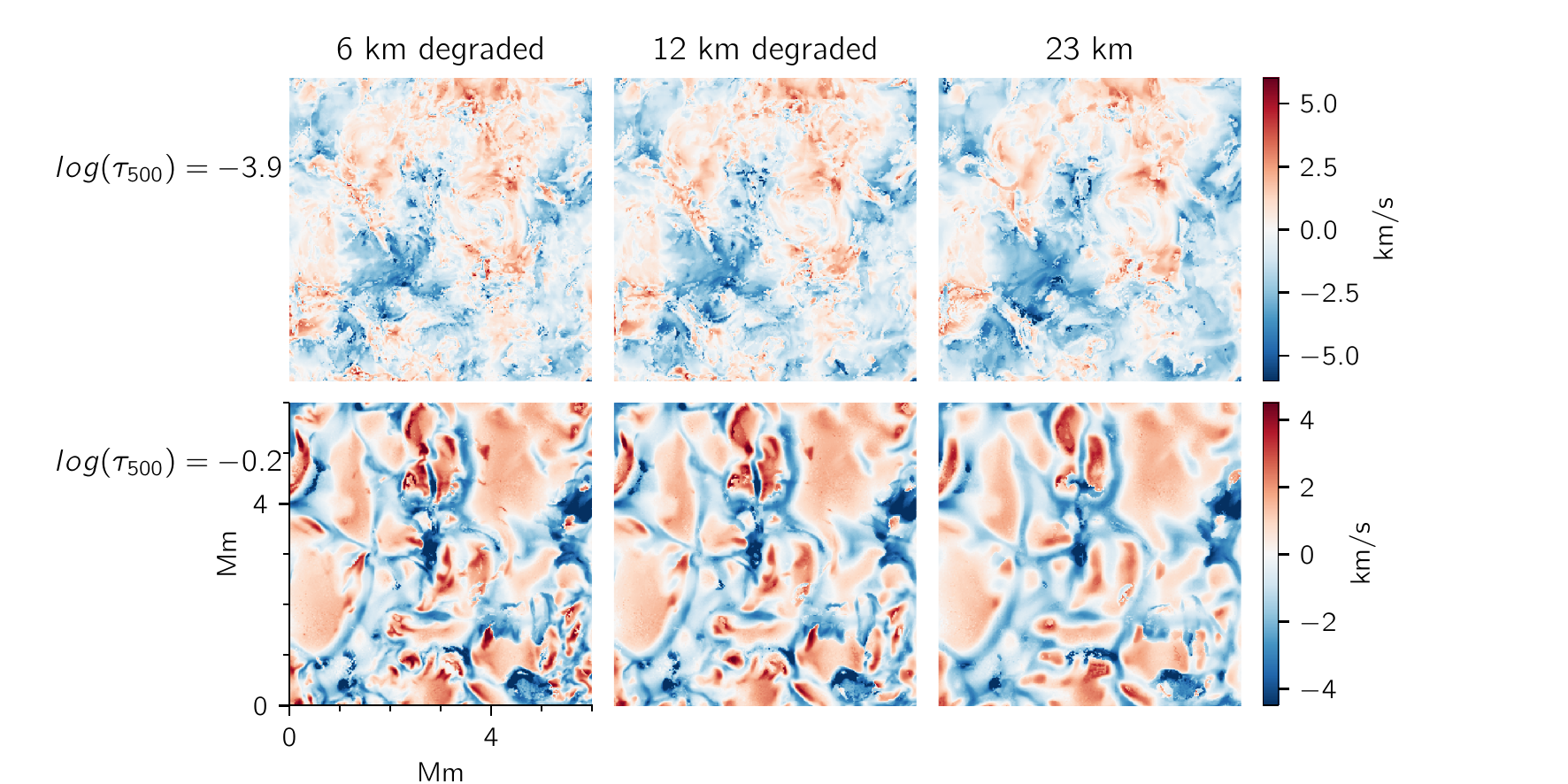}
			\caption{Retrieved line-of-sight velocities from the inversions of the (degraded) spectra, taken at the heights $\log(\tau_{500}) = -3.9$ (top) and $log(\tau_{500}) = -0.2$ (bottom), corresponding to the lower chromosphere and the photosphere respectively.}
			\label{fig:inv_vel}
	\end{figure*}
	
	\begin{figure*}
	\centering
		\includegraphics[width=18cm]{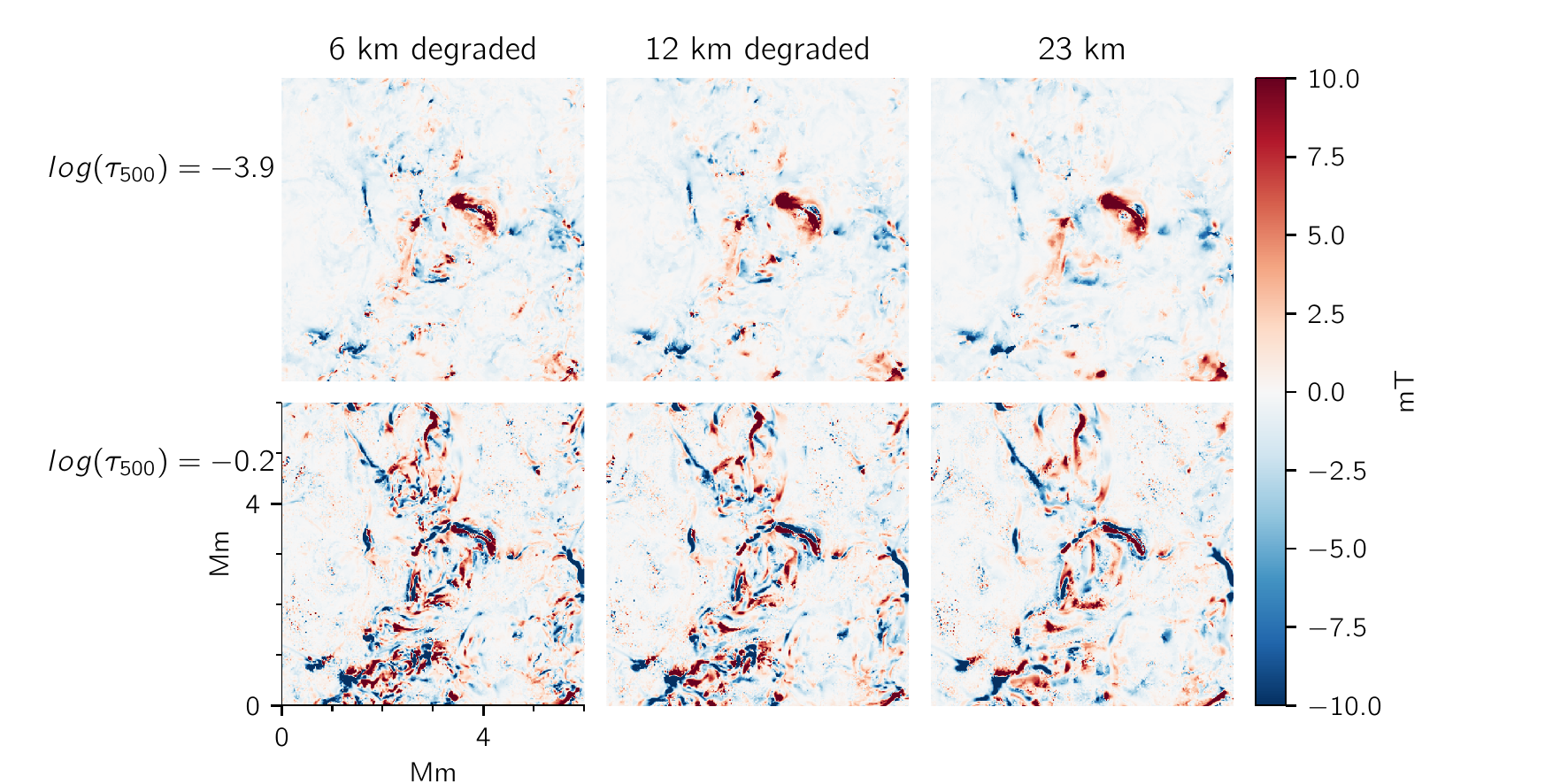}
			\caption{Retrieved line-of-sight magnetic field strengths from the inversions of the (degraded) spectra, taken at the heights $\log(\tau_{500}) = -3.9$ (top) and $log(\tau_{500}) = -0.2$ (bottom), corresponding to the lower chromosphere and the photosphere respectively.}
			\label{fig:inv_mag}
	\end{figure*}
	
	\begin{figure*}
	\centering
		\includegraphics[width=18cm]{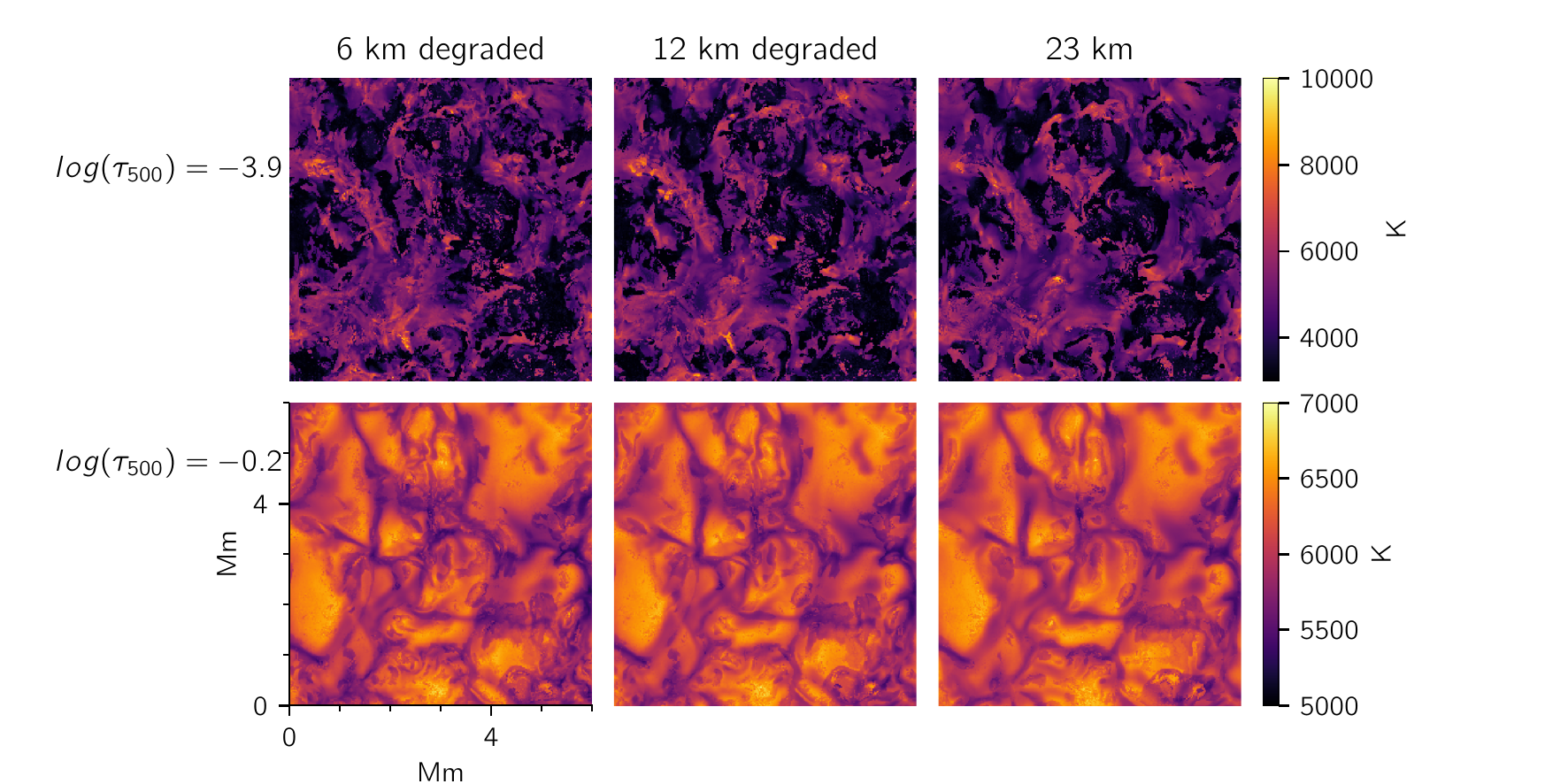}
			\caption{Retrieved temperatures from the inversions of the (degraded) spectra, taken at the heights $\log(\tau_{500}) = -3.9$ (top) and $log(\tau_{500}) = -0.2$ (bottom), corresponding to the lower chromosphere and the photosphere respectively.}
			\label{fig:inv_temp}
	\end{figure*}
	
	\begin{figure*}
	\centering
		\includegraphics[width=18cm]{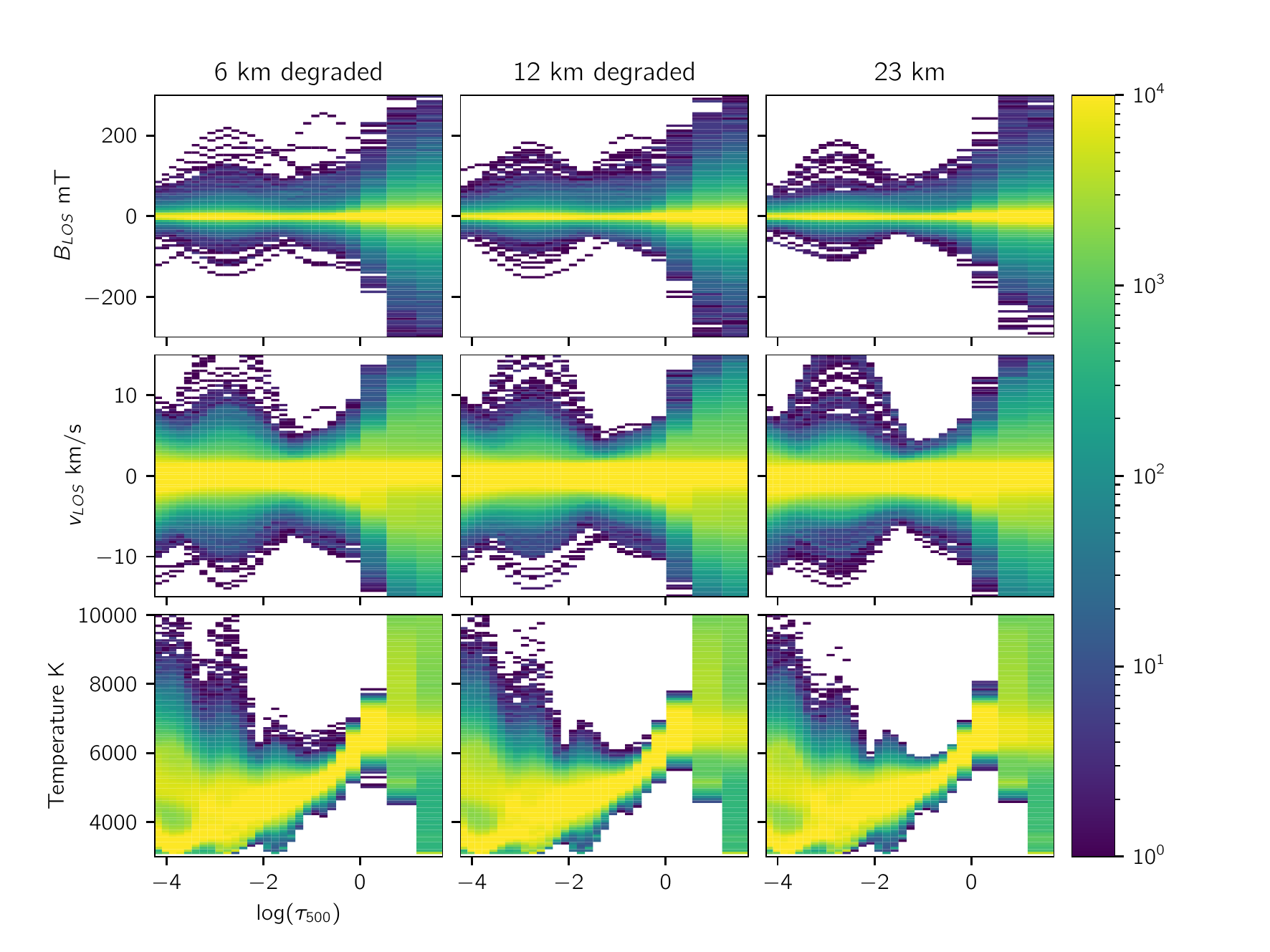}
			\caption{Distribution (number of pixels) of retrieved line-of-sight velocities and magnetic field strengths, as well as temperature, for the inversions of the (degraded) spectra, taken at each considered $\log(\tau_{500})$ grid point of the inverted atmospheres. The color bar is logarithmic in the number of pixels, and saturated at $10^4$.}
			\label{fig:inv_hist}
	\end{figure*}
	
	\begin{figure}
		\centering
			\includegraphics[width=9cm]{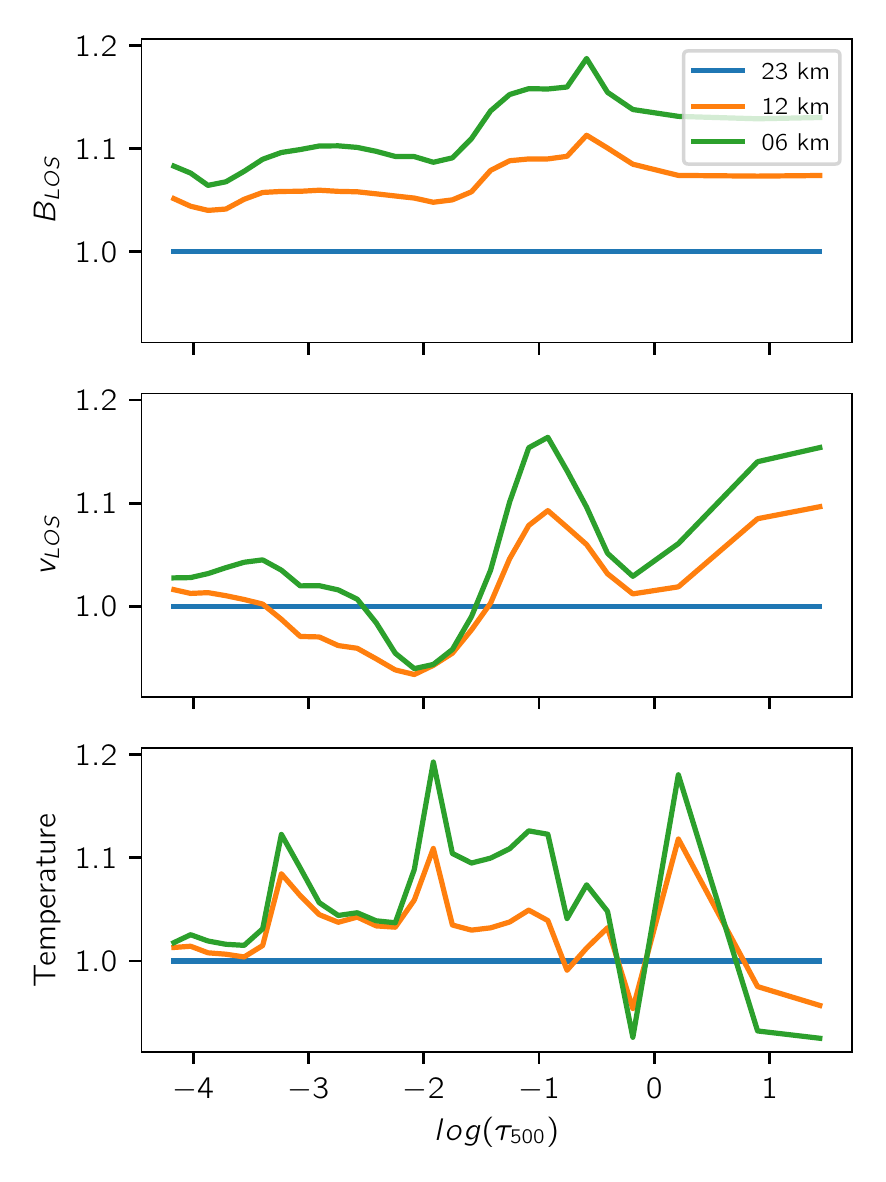}
				\caption{Median absolute difference between inversion and simulation quantities taken at each considered $\log(\tau_{500})$ grid point of the inverted atmospheres. Normalized to the median absolute difference of the 23 km inversion and the 23 km simulation}
				\label{fig:inv_relative_diff}
		\end{figure}

\subsection{Effects on inversions}
\label{subsec:inversion_results}

We now turn our attention to the effects on quantities inferred from inversions. Our primary goal was not to perform the best possible inversions that most accurately reproduce the quantities from the simulations. Instead, we focus on the differential effects that arise from using spectra with varying amounts of sub-resolution effects. It may be possible to tweak and fine tune the parameters of the inversion to obtain a better overall fit, but we believe our inversion results are good enough for our purposes.

We show vertical cuts of inverted line-of-sight velocities, line-of-sight magnetic field strengths, and temperatures at two heights, respectively in Figs.~\ref{fig:inv_vel}, \ref{fig:inv_mag}, and \ref{fig:inv_temp}. These were taken at $\log(\tau_{500} )= -3.9$ (top) and $\log(\tau_{500}) = -0.2$ (bottom). These heights are not directly comparable to neither the previously shown simulation slices, nor to the formation heights of the computed spectral lines. They were chosen to show the terrain in the low chromosphere and photosphere, respectively. In particular, the higher point is slightly below the mean $\tau = 1$ height of the \ion{Ca}{ii} line core. After this point, we are beginning to move outside the formation region of our two spectral lines, and therefore the inversions are not sensitive to these regions. For that reason, in our analysis we disregard regions with $\log(\tau_{500} )\lesssim -4 $. 

As before, we see more fine-structure with increasing resolution in all cases, also for the temperature maps. From the 23 km to 12 km case we also see the ``hotspots'' of increased extreme values in line-of-sight velocities and magnetic field strengths. Less pronounced, but still visible by inspection, is the same tendency comparing the 12 km inversion to the 6 km inversion. In particular the photospheric line-of-sight velocities, but also the line-of-sight magnetic field strengths and (most faintly) the chromospheric line-of-sight velocities show these localized areas of increased extremes. For the temperatures, the tendency of increased extremes still holds going from 23 km to 12 km, but now we also note that some ``hotspots'' disappear from the maps along the way, i.e. some areas of extreme temperatures become smoother at higher resolution. For instance, the bright (yellow) spots seen at $(x,y)\approx(2,2)$ Mm in both 23~km panels, or $(x,y)\approx(3,5)$ Mm in the lower panel of Fig. \ref{fig:inv_temp}. Interestingly, from the 12 km case to the 6 km case the variation in extreme areas does not appear to follow any particular trend, at least on visual inspection. This, together with the less pronounced variation in fine structure, suggests a closer convergence in the retrieved temperatures than in the other atmospheric quantities. 

The new element brought in by the inversions is the presence of obvious artifacts, both salt-and-pepper-like pixels that stand out as well as larger regions of more continuous over- and under-estimations. The latter are especially poignant around the regions of strong line-of-sight magnetic field strengths, where artificial regions of opposing polarity (not present in either the simulations or the simple inferences discussed earlier) are imposed around the real fields. These smooth regions are particularly intriguing when one considers that the STiC inversion code treats each column independently in a 1.5D fashion. These spurious regions are more frequent for 6~km than 12~km, suggesting that sub-resolution effects have a systematic impact that negatively affects the accuracy of inversions. This could be interpreted as the sub-resolution effects making it more difficult to reproduce the degraded spectra with a 1D plane-parallel atmosphere, since the individual atmospheric columns making up the pixel to be inverted span a broader range of conditions.

We show the distributions (number of pixels, color-coded on a logarithmic scale) of the inverted quantities versus their $\log(\tau_{500})$ in Fig. \ref{fig:inv_hist}. The salient features of these images are the ranges and numbers of the uttermost extremes, and, most importantly, the steepness of the color-gradient for the middling values. From this figure it is apparent that more extreme line-of-sight velocities can be found for increasing resolution at all considered heights. Regarding the line-of-sight magnetic field strengths, one can see a wider distribution for the 6 km and 12 km case versus the 23 km case. Comparing the 6 km inversion with the 12 km inversion, the differences do not seem as large; but importantly there seems to be more negative line-of-sight magnetic fields present throughout the considered heights, in agreement with the increased amount of (artificial) negative field in Fig. \ref{fig:inv_mag}. For the temperatures, the differences between 6 km and 12 km appear particularly small. From the distributions alone, it is not apparent that the 6 km inversion produces worse artifacts than the 12 km, but comparing the photospheric inversion results with both the simulation variables and the inferences from the spectra shows that one of the effects of increasing the underlying spatial resolution is that inversions seem to converge on incorrect solutions, which introduce erroneous fields.

To quantify the reliability of the inversions with varying amounts of sub-resolution effects, we looked at the variation of the median absolute difference between the atmospheric quantities inferred from inversions and the corresponding simulation variables as a function of $\log(\tau_{500})$. We interpolated the line-of-sight magnetic fields, velocities and the temperature for each of the three simulations to the $\log(\tau_{500})$ grid points of the inversions. We subsequently performed the same spatial degradation on these interpolated quantities as we did to the synthetic profiles. For each $\log(\tau_{500})$ iso-surface we computed the median absolute difference between the inversion and the corresponding (degraded) simulation, normalized to the values for the 23~km runs. These median differences therefore give a statistical measure of how accurate inversions are at different depths, compared to a baseline set by the 23~km simulation, which has no sub-resolution effects. We show the results of this procedure in Fig. \ref{fig:inv_relative_diff}.

We find that sub-resolution negatively affects the reliability of inversions at nearly all optical depths, by about 10\%. Compared to simulation variables degraded in the same way as the spectra used for inversions, the inverted variables are consistently worse when compared to the case with no sub-resolution, the 23~km run. There are some heights where the inverted and degraded atmospheres do match better in line-of-sight velocities and temperature for the higher-resolution cases. However, on the whole, the inclusion of sub-resolution imprints on the spectra to be inverted seems to reduce the accuracy of the inferred atmosphere by about 5--20\%, typically around 10\%. All three quantities, temperature, line-of-sight velocity and line-of-sight magnetic field strength are affected by similar amounts. The results for the 6~km simulation are worse than for the 12~km simulation, indicating that with increased sub-resolution, the inversions become less reliable. Encouragingly, the difference from 6~km to 12~km is much smaller than the difference from 12~km to 23~km, which suggests that we are converging on an upper limit. Although one should be careful since we have only two data points, it is worth noting that the results seem robust because the differences between 6~km and 12~km are consistent for different variables and across different depths, and that the median, taken over $512\times512$ points for each value of $\log(\tau_{500})$ will neglect the more extreme ``salt and pepper'' cases where the inversions struggled. That the results seem to vary a lot less from 6~km to 12~km is also consistent with the idea that adding more and more sub-resolution effects will have a diminishing effect on the spectra, since much higher resolution will be strongly suppressed by the spatial convolution, and result in subtler and subtler effects on the spectra.

\section{Discussion and Conclusions}
\label{sec:conclusion}

Adopting a reference numerical grid size of 23~$\mathrm{km}\,\mathrm{pix}^{-1}$, we analyzed full Stokes spectra from three simulations with pixel sizes of 23, 12, and 6~$\mathrm{km}\,\mathrm{pix}^{-1}$, with the spectra from the latter two degraded to the reference resolution. The purpose was to build spectra with varying degrees of sub-resolution imprints. For the 23~km run they are at the native resolution (no sub-resolution), for 12~km at half the native resolution, and for 6~km a quarter of the native resolution. Our original question was to find how much these sub-resolution effects impact inferred quantities such as line-of-sight velocities and magnetic fields.

Our results show that simulations with different numerical resolutions do produce different results, and that these differences persist even after a spatial degradation of the emergent spectra. While the large scale structures and tendencies remain the same between the simulations that have been considered here, the higher-resolution simulations produce fine structure beyond that of the lower-resolution simulations, and this fine structure still remains after applying a Gaussian convolution and downsampling. Furthermore, the atmospheric quantities inferred from the higher-resolution simulations show a higher occurrence of more extreme values when compared to the lower resolution simulations. From the distributions of the inferred quantities, we find that the trend of the extreme values increasing in tandem with the resolution clearly persists even up to the 6 km resolution in photospheric line-of-sight velocities and chromospheric line-of-sight magnetic field strengths. The same trend is less clear when it comes to the histograms of photospheric line-of-sight magnetic field strengths, however, as the 12 km and 6 km spectra give rise to comparable maximum values of the field. Still, even in this case, one can clearly see more fine-structure in the inferred photospheric line-of-sight magnetic field maps for the 6 km simulation when compared to the 12 km simulation.

When it comes to inversions, our results show a consistent and clear trend: sub-resolution leads to a systematic error. This error is not much, about 10\%, but is present at all heights and in the three variables analyzed here: line-of-sight velocity, line-of-sight magnetic field strength, and temperature. Our analysis is limited to resolutions that are typical of 1-m class telescopes, but with the power-law scaling of small-structure in the solar surface \citepads{2012ApJ...756L..27A}, similar effects are likely to be present even with the next generation of 4-m telescopes.

It is important to be aware of the limitations of this work. Given how the simulations were devised, and because we wanted to compare on a pixel-by-pixel basis, we were limited in having only three resolutions, and we used only one snapshot for each resolution. To have a larger range of resolutions based on the same parent simulation would need additional relaxation, with the consequence that changes would develop in the evolution of the atmosphere and the fields of view would be different. In that case, only a statistical study would be possible. We note also that given how the simulations were built, the vertical resolution was not the same between the 23~km resolution and the other two, which had twice as many points. It is possible that some of the differences in the inversions arise from these changes in the vertical resolution and not solely because of the changes in horizontal resolution. Our spatial degradation was purely Gaussian, not following an Airy disk diffraction nor a point spread function (PSF) that could also include atmospheric seeing. Since the geometries of different instruments lead to different PSFs, we chose a Gaussian convolution since it is more neutral, and we believe the differences between an Airy pattern and the Gaussian function are negligible for our purposes, especially since the FWHM used are only 2 or 4 pixels. We also did not include any spectral degradation, since our focus was on spatial resolution. Finally, we ran the inversions with a simple setup, which could have been better optimised. A more careful tweaking of the parameters could have led to better performance, but having the ground truth (the simulations) means that any adjustments using that knowledge are not neutral when compared to an inversion of real observations. This justified having a simple setup, also because our focus was on the differential effect of resolution in the inversions, not on the absolute performance of the inversions, which are also subject to other sources of systematic errors.

To model phenomena extending to a given physical scale one must choose a suitable numerical grid size. Numerical grid sizes are not directly comparable with observational resolution or pixel sizes. While a deep analysis of these differences is outside the scope of this work, we find that given the numerical diffusion present in these Bifrost experiments, a grid size of 23~$\mathrm{km}\,\mathrm{pix}^{-1}$ is roughly equivalent to an observational resolution of about 100~$\mathrm{km}\,\mathrm{pix}^{-1}$. Notwithstanding the differences between numerical and observational pixel sizes, we argue that to mitigate for sub-resolution effects it is best to aim for a numerical grid size that is about half that of the target value. If one assumes our equivalency of 100~km on the sun with a numerical grid of 23~km, our results indicate that a grid of 12~km will go a long way at mitigating small-scale structure that still leaks into inferred quantities. Going for a grid size four times smaller than the target value will give a more accurate result, but with diminishing returns, especially taking into account the much larger computational cost.

\begin{acknowledgements}
The authors would like to thank Jorrit Leenaarts, Rob Rutten and Oskar Steiner for giving helpful comments, questions and suggestions at preliminary talks about the contents of this manuscript, Åke Nordlund for useful comments regarding the effects of numerical diffusion, as well as Jaime de la Cruz Rodr{\'\i}gues for providing valuable help with setting up the STiC-inversions. We would also like to thank the anonymous referee for constructive and thorough feedback leading to several improvements of this manuscript. This work has been supported by the Research Council of Norway through its Centers of Excellence scheme, project number 262622. Computational resources have been provided by UNINETT Sigma2 - the National Infrastructure for High Performance Computing and Data Storage in Norway.
\end{acknowledgements}

\bibliography{Refs.bib}
\end{document}